\documentclass[
    ,final            % use final for the camera ready runs
  ]
  {aipproc}

\layoutstyle{6x9}

\begin{document}

\title{VERITAS Observations of M\,87 in 2011/2012}

%http://www.aip.org/pacs/index.html


\author{M.~Beilicke}{address={Department of Physics and MCSS, Washington
University, St. Louis, MO, USA}, email={beilicke@physics.wustl.edu}}

\author{the VERITAS
Collaboration}{address={\url{http://veritas.sao.arizona.edu/}}}

\begin{abstract}

The giant radio galaxy M\,87 is located at a distance of $16.7 \,
\rm{Mpc}$ and harbors a super-massive black hole (6 billion solar
masses) in its center. M\,87 is one of just three radio galaxies known
to emit TeV $\gamma$-rays. The structure of its relativistic plasma jet,
which is not pointing towards our line of sight, is spatially resolved
in X-ray (Chandra), optical and radio (VLA/VLBA) observations. The
mechanishm and location of the TeV emitting region is one of the least
understood aspects of AGN. In spring 2008 and 2010, the three TeV
observatories VERITAS, MAGIC and H.E.S.S. detected two major TeV flares
in coordinated observations. Simultaneous high-resolution observations
at other wavelengths~-- radio (2008) and X-rays (2008/2010)~-- gave
evidence that one of the TeV flares was related to an event in the core
region; however, no common/repeated patterns could be identified so far.
VERITAS continued to monitor M\,87 in 2011/2012. The results of these
observations are presented.

\end{abstract}

\maketitle

%%%%%%%%%%%%%%%%%%%%%%%%%%%%%%%%%%%%%%%%%%%%
%%%%%%%%%%%%%%%%%%%%%%%%%%%%%%%%%%%%%%%%%%%%
%% MAINMATTER
%%%%%%%%%%%%%%%%%%%%%%%%%%%%%%%%%%%%%%%%%%%%
%%%%%%%%%%%%%%%%%%%%%%%%%%%%%%%%%%%%%%%%%%%%

%% ############################################################
%% ############ Introduction
%% ############################################################
\section{Introduction}

Due to its super-massive black hole and its close proximity of $16.7 \,
\rm{Mpc}$ structures of the relativistic plasma jet in M\,87 can be
resolved in radio to X-ray observations \cite{Wil02}. This makes M\,87 a
unique AGN that potentially allows to spatially constrain the location
of the TeV $\gamma$-ray emission (radio/X-ray vs.~TeV flux correlation
studies). Such a result would provide important input for AGN
unification theories and plasma jet physics.

M\,87 is now a well established TeV $\gamma$-ray source
\cite{Aha03,Goe06,Aha06,Acc08} with observed TeV flux variability on
time scales of days \cite{Aha06,Alb08,Acc09}. The variability strongly
constrains the size of the emission region to the order of ten
Schwarzschild radii. This rules out interpretations related to large
scale emission from dark matter annihilation in M\,87 \cite{Bal99} or
cosmic ray interaction \cite{Pfr03}. Leptonic \cite{Geo05,Len08,Cui12}
and hadronic \cite{Rei04} jet emission models are discussed in the
literature as well as different potential locations of the TeV
$\gamma$-ray emission: the nucleus \cite{Ner07}, the inner jet
\cite{Geo05,Rei04} or the (large scale) jet \cite{Che07,Tav08,Hrd11}.

Since 2007 teV $\gamma$-ray observations of M\,87 have been closely
coordinated between VERITAS, MAGIC, and H.E.S.S. This drastically
improved the chances of detecting active flux states and substantially
increased the scientific outcome of the observations. 
Target-of-opportunity (ToO) observations were proposed at radio (VLBA)
to X-ray (Chandra) energies. So far, three TeV $\gamma$-ray flares have
been detected in the past 13 years of M\,87 observations:

{\bf 2005:} H.E.S.S. observations in 2004/2005 confirmed M\,87 as a TeV
$\gamma$-ray source. Surprisingly, flux variability was found on time
scales of days \cite{Aha06}. The year 2005 also marked the maximum of an
radio/optical/X-ray high state of the innermost knot in the jet, HST-1,
located roughly $60 \, \rm{pc}$ away from the core \cite{Har06}. Since
this high state lasted for almost one year~-- with correspondingly slow
time scales of flux changes~-- it is not clear whether it was related to
the TeV flare.

{\bf 2008:} A strong TeV $\gamma$-ray flare was detected in joint
VERITAS, MAGIC, and H.E.S.S. observations in 2008 confirming the
day-scale flux variability \cite{Acc10,Aha06,Alb08}. The flare marked
the onset of a rise in the $43 \, \rm{GHz}$ radio flux (VLBA) from the
nucleus. The VLBA observations resolve the nucleus with very high
precision of $30 \times 60$ Schwarzschild radii; the radio nucleus is
believed to be coincident with the black hole position \cite{Had11}.
Combining the radio and TeV data therefore gave first experimental
evidence that the TeV emission actually originates from the close
vicinity of the black hole \cite{Acc09}.

\begin{figure}[t]

\includegraphics[width=0.49\textwidth]{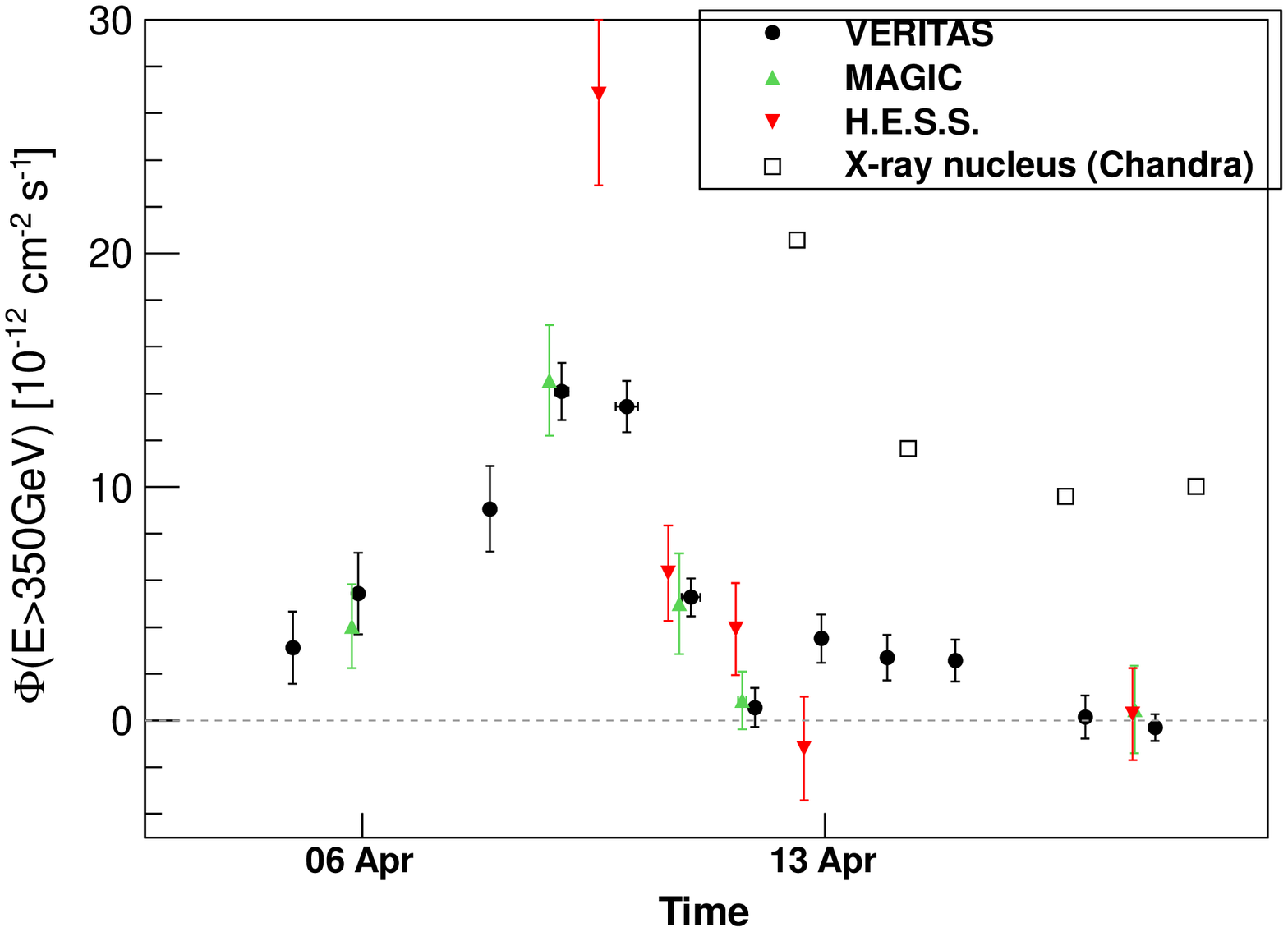}
\hfill
\includegraphics[width=0.49\textwidth]{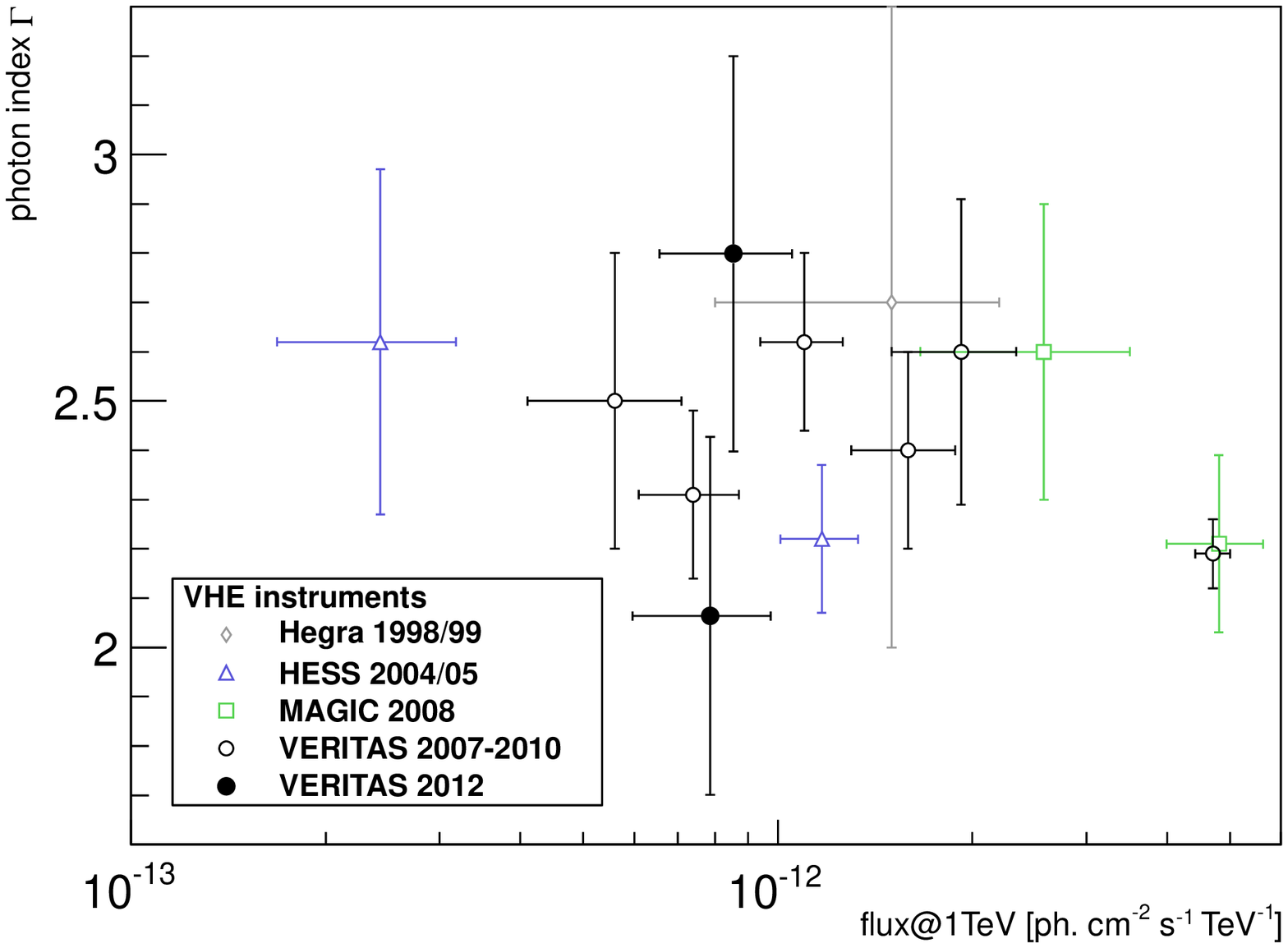}

\caption{\label{fig:General} {\bf Left:} The night-by-night light curve
of the 2010 TeV $\gamma$-ray flare (VERITAS, MAGIC and H.E.S.S.; data
taken from \cite{Abr12}). Also shown is the Chandra X-ray light curve
(arbitrary units) taken from \cite{Har11}. {\bf Right:} TeV flux
normalization vs. photon index $\Gamma$. A spectral hardening for high
fluxes is not yet significant.  Data are taken from
\cite{Aha03,Aha06,Acc08,Alb08,Acc10} as well as the data set presented
in this paper.}

\end{figure}

{\bf 2010:} The strongest TeV $\gamma$-ray flare so far was detected by
VERITAS and MAGIC in 2010 (Fig.~\ref{fig:General}, left), also
triggering follow-up observations by H.E.S.S.~\cite{Ali12,Abr12}. Radio,
optical, and X-ray observations were triggered in addition to the
accompanying Fermi/LAT observations of the source. Although the flux of
the radio nucleus remained at the base level the weeks following the TeV
flare~-- not confirming the 2008 finding~-- two interesting observations
were made in the contemporaneous multi-wavelengths (MWL) monitoring: (i)
the Chandra observations (starting $\sim$$4 \, \rm{d}$ after the peak of
the TeV flare) showed the second highest flux measured from the nucleus
so far \cite{Har11} (Fig.~\ref{fig:General}, left); (ii) a new radio
feature was identified in the knot HST-1 following the flare
\cite{Gir12}.

%-------------
\begin{figure}[t]

\includegraphics[width=0.49\textwidth]{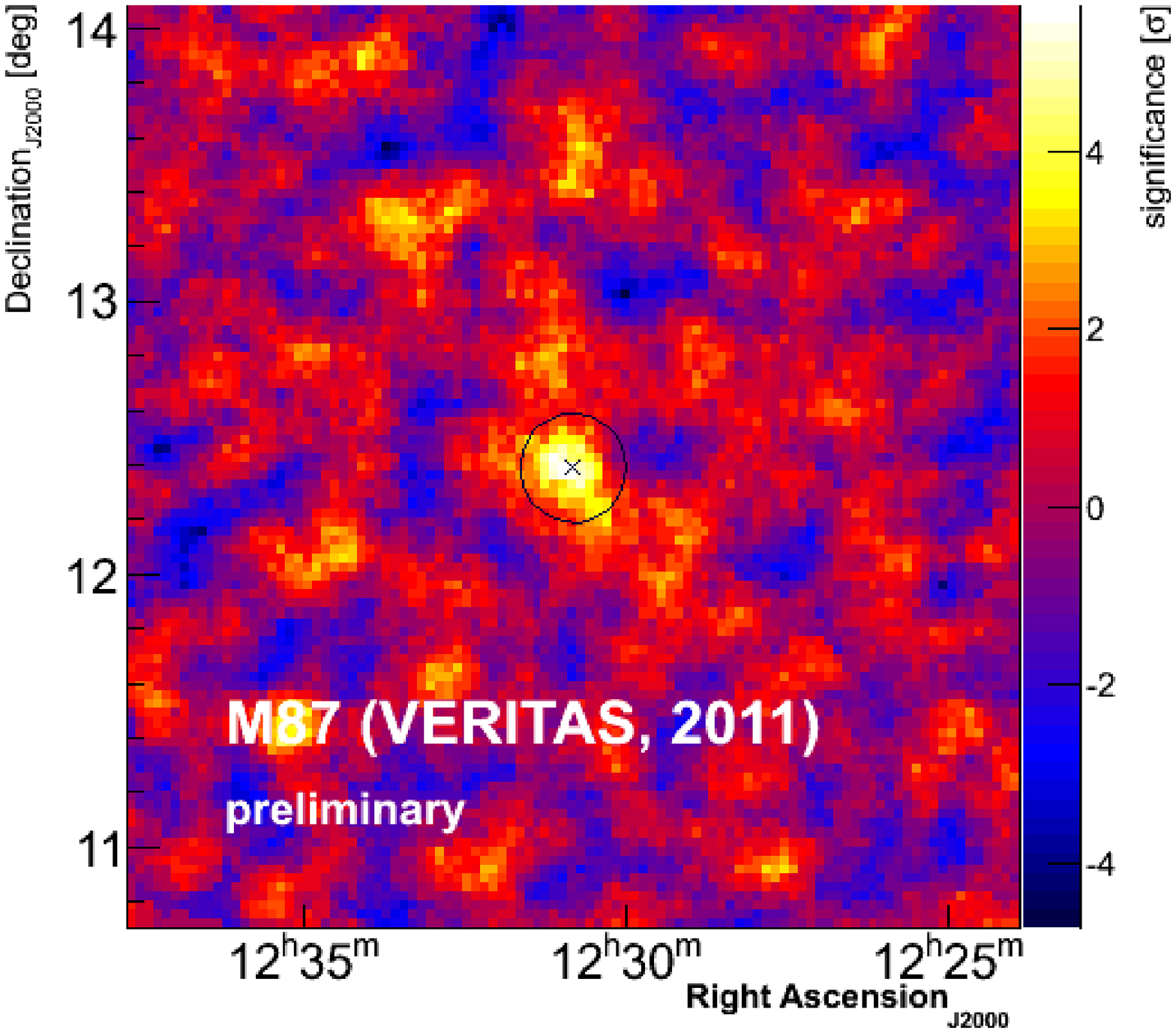}
\hfill
\includegraphics[width=0.49\textwidth]{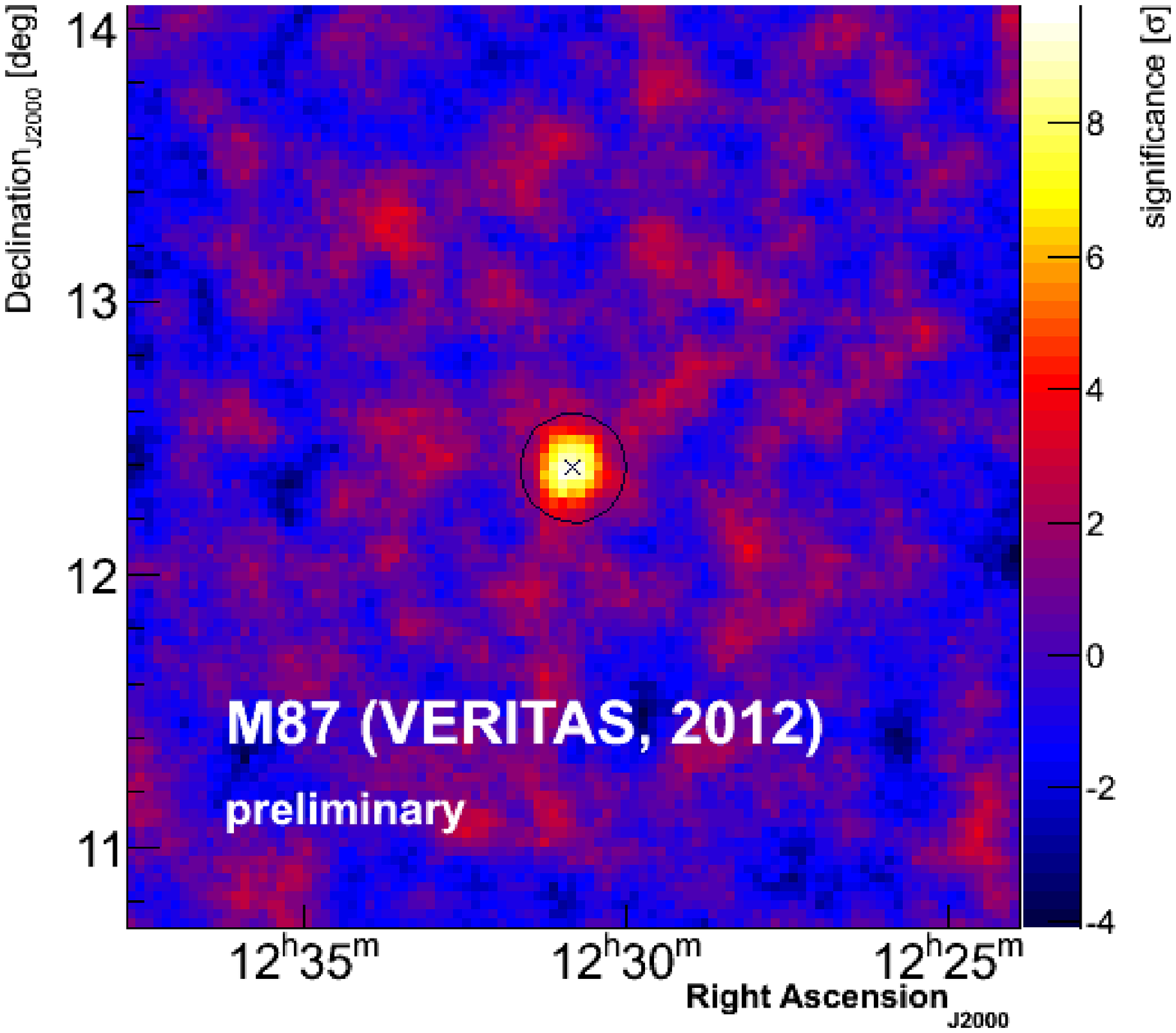}

\caption{\label{fig:Skymap} M\,87 sky maps (smoothed excess
significances, ring background) measured by VERITAS in 2011 (left) and
2012 (right).}

\end{figure}

%% ############################################################
%% ############ VERITAS observations of M\,87 in 2011/2012
%% ############################################################
\section{VERITAS observations of M\,87 in 2011/2012}

\begin{figure}[t]

\includegraphics[width=0.49\textwidth]{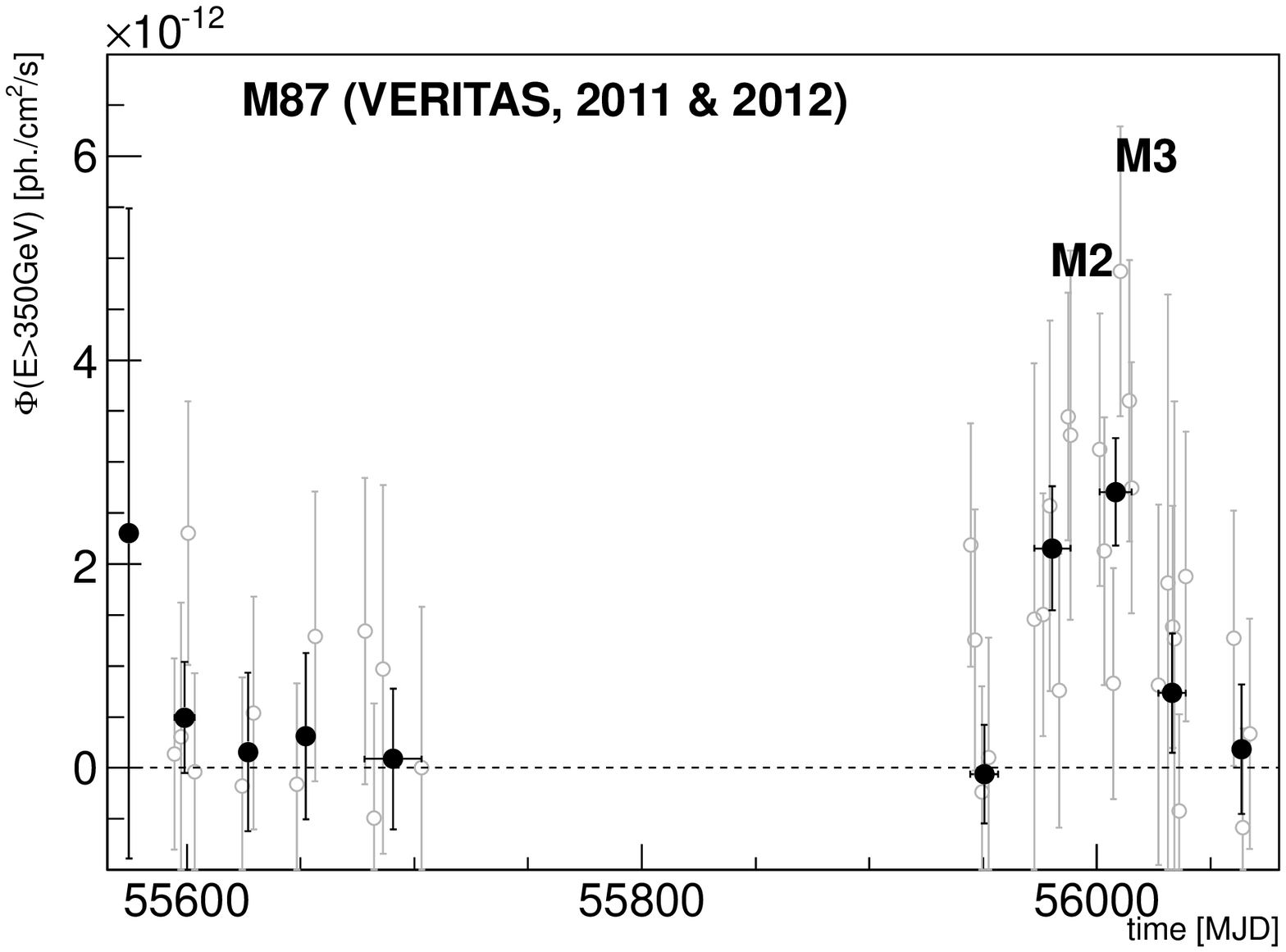}
\hfill
\includegraphics[width=0.49\textwidth]{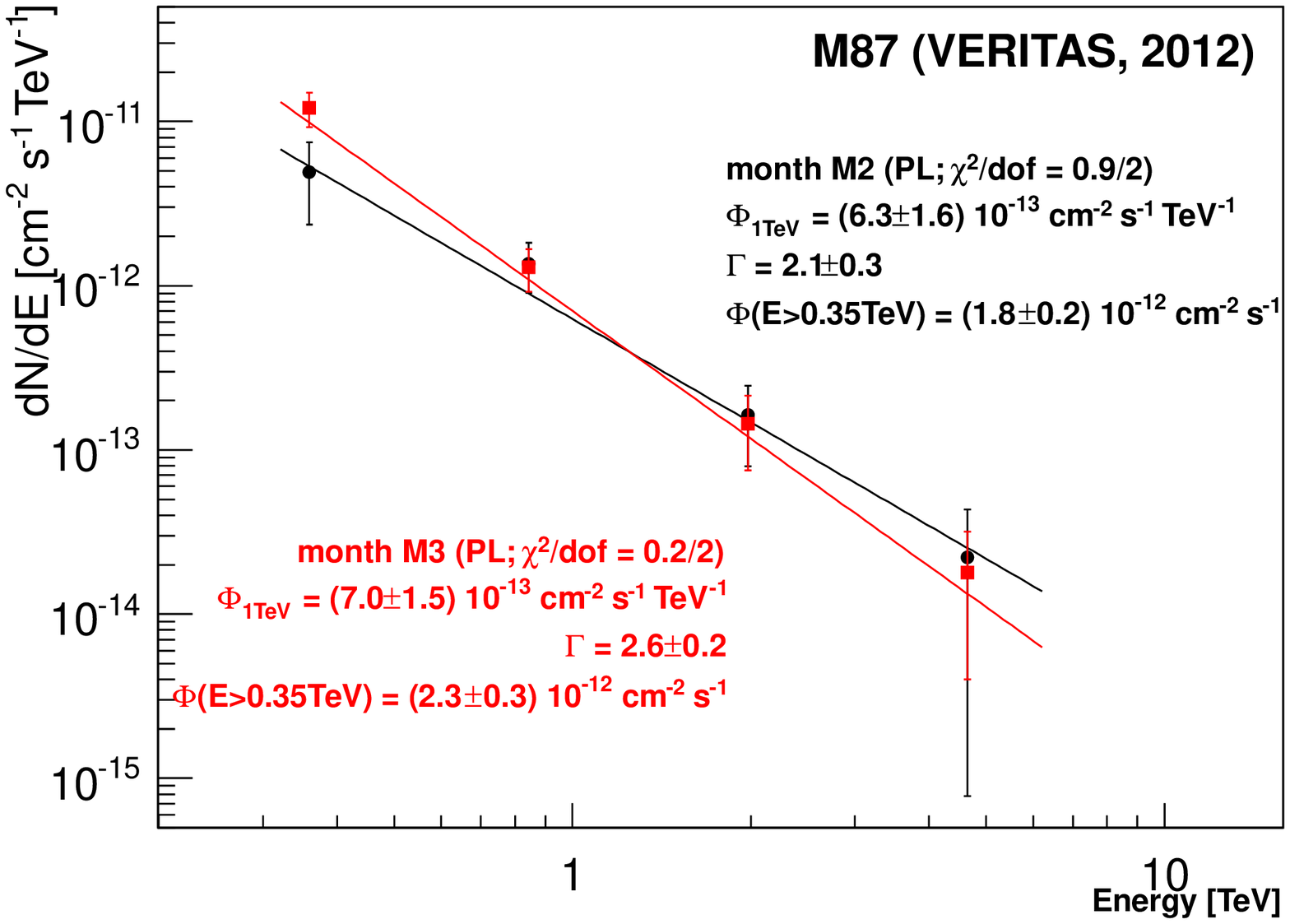}

\caption{\label{fig:SED} {\bf Left:} VERITAS light curve of M\,87 in
2011/2012 (gray: nightly fluxes; black: monthly fluxes). Elevated flux
levels were measured in 2012 during the months M\,2 and M\,3. {\bf
Right:} Energy spectra derived for data sets M\,2 and M\,3. The spectra
were fitted by power laws.}

\end{figure}

VERITAS continued monitoring M\,87 in 2011/2012. The observations were
coordinated with MAGIC, H.E.S.S., and the MWL partners of the previous
campaigns. A total of $17 \, \rm{hrs}$ of good quality data were
accumulated by VERITAS in 2011. Only a weak signal at the level of 5
std.~dev.~was detected, see Fig.~\ref{fig:Skymap}, left. The 2012
observations resulted in a total of $29 \, \rm{hrs}$ of data. M\,87 is
detected at the level of 9 std.~dev., see Fig.~\ref{fig:Skymap}, right.
The light curve of the two years is shown in Fig.~\ref{fig:SED}, left,
and indicates variability on time scales of weeks (but not days) in
2012: two months (M\,2 and M\,3) with clearly elevated flux are
identified. The energy spectra derived for these two months are shown in
Fig.~\ref{fig:SED}, right, and are found to be compatible with each
other. Figure~\ref{fig:General} shows the photon index $\Gamma$ vs.~the
TeV flux normalization of these spectra as well as for archival data.
Nightly flux levels that would allow to trigger the MWL observations
were not reached in 2011/2012.

%-------------

%% ############################################################
%% ############ Summary and Conclusion
%% ############################################################
\section{Summary and conclusion}

M\,87 was monitored by VERITAS in 2011/12 for $36 \, \rm{hrs}$ in
coordination with MAGIC and H.E.S.S. While the VERITAS data of 2011 and
2012 did not show bright flares, the monthly variation in TeV flux hint
that the underlying quiescent emission also evolves over much longer
time scales than the day-scale rapid flares seen in 2008 and 2010. This
makes long term monitoring of M 87 in TeV and other wavebands
scientifically very fruitful.

Although promising indications were found in TeV/MWL observation of
previous years the location and mechanisms of the TeV $\gamma$-ray
emission are still uncertain. However, due to its proximity and jet
viewing angle, M\,87 remains a unique laboratory to study the connection
between jet formation physics and TeV $\gamma$-ray emission
\cite{Bro11,Dex12,Lev11}. Coordinated TeV $\gamma$-ray observations
accompanied by contemporaneous observations at radio (high-resolution),
X-ray and Fermi energies remain one of the most promising approach to
unravel the location and mechanism of TeV $\gamma$-ray emission in AGN.

%-------------

%%%%%%%%%%%%%%%%%%%%%%%%%%%%%%%%%%%%%%%%%%%%%%%%
%% Acknowledgments
%%%%%%%%%%%%%%%%%%%%%%%%%%%%%%%%%%%%%%%%%%%%%%%%

\begin{theacknowledgments}

This research is supported by grants from the U.S. Department of Energy
Office of Science, the U.S. National Science Foundation and the
Smithsonian Institution, by NSERC in Canada, by Science Foundation
Ireland (SFI 10/RFP/AST2748) and by STFC in the U.K. We acknowledge the
excellent work of the technical support staff at the Fred Lawrence
Whipple Observatory and at the collaborating institutions in the
construction and operation of the instrument.

\end{theacknowledgments}


\bibliographystyle{aipproc}   % if natbib is available

%\bibliography{sample}

%%%%%%%%%%%%%%%%%%%%%%%%%%%%%%%%%%%%%%%%%%%
%% The Bibliography
%%%%%%%%%%%%%%%%%%%%%%%%%%%%%%%%%%%%%%%%%%%

\end{document}